Malte Möser*, Kyle Soska, Ethan Heilman, Kevin Lee, Henry Heffan, Shashvat Srivastava, Kyle Hogan, Jason Hennessey, Andrew Miller, Arvind Narayanan, and Nicolas Christin

# An Empirical Analysis of Traceability in the Monero Blockchain

**Abstract:** Monero is a privacy-centric cryptocurrency that allows users to obscure their transactions by including chaff coins, called "mixins," along with the actual coins they spend. In this paper, we empirically evaluate two weaknesses in Monero's mixin sampling strategy. First, about 62% of transaction inputs with one or more mixins are vulnerable to "chain-reaction" analysis — that is, the real input can be deduced by elimination. Second, Monero mixins are sampled in such a way that they can be easily distinguished from the real coins by their age distribution; in short, the real input is usually the "newest" input. We estimate that this heuristic can be used to guess the real input with 80% accuracy over all transactions with 1 or more mixins. Next, we turn to the Monero ecosystem and study the importance of mining pools and the former anonymous marketplace AlphaBay on the transaction volume. We find that after removing mining pool activity, there remains a large amount of potentially privacy-sensitive transactions that are affected by these weaknesses. We propose and evaluate two countermeasures that can improve the privacy of future transactions.

**\*Corresponding Author: Malte Möser:** Princeton University, E-mail: mmoeser@princeton.edu
**Kyle Soska:** Carnegie Mellon University, E-mail: ksoska@cmu.edu
**Ethan Heilman:** Boston University, E-mail: ethan.r.heilman@gmail.com
**Kevin Lee:** University of Illinois at Urbana-Champaign, E-mail: klee160@illinois.edu
**Henry Heffan:** Brookline High School, E-mail: henry.heffan@gmail.com
**Shashvat Srivastava:** Massachusetts Academy of Math and Science at WPI, E-mail: ssrivastava2@wpi.edu
**Kyle Hogan:** Massachusetts Institute of Technology, E-mail: klhogan@mit.edu
**Jason Hennessey:** Boston University, E-mail: henn@bu.edu
**Andrew Miller:** University of Illinois at Urbana-Champaign, E-mail: soc1024@illinois.edu
**Arvind Narayanan:** Princeton University, E-mail: arvindn@cs.princeton.edu
**Nicolas Christin:** Carnegie Mellon University, E-mail: nicolasc@cmu.edu

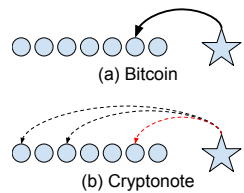

**Fig. 1.** Transactions and tracing in Bitcoin and Cryptonote. Consider a new transaction (the star) which spends an available coin (the second circle from the right). In Bitcoin (a), each transaction input explicitly identifies the coin being spent, thus forming a linkage graph. In the Cryptonote protocol (b), each transaction input identifies a set of coins, including the real coin along with several chaff coins called "mixins." However, many mixins can ruled out by deduction (Section 3); furthermore, the real input is usually the "newest" one (Section 4).

## 1 Introduction

Monero is a leading privacy-centric cryptocurrency based on the Cryptonote protocol. As of November 2017 it is one of the most popular cryptocurrencies at a market capitalization of USD 1.5B. While Bitcoin, the first and currently largest cryptocurrency, explicitly identifies which coin in the transaction graph is being spent, Cryptonote allows users to obscure the transaction graph by including chaff transaction inputs called "mixins" (this is visualized in Figure 1).

As a result, Monero has attracted the attention of users requiring privacy guarantees superior to those Bitcoin provides. Some of the most publicized uses are illicit (for instance, the former online anonymous marketplace AlphaBay accepted Monero as a payment instrument), but we estimate that illicit use accounts for at most 25% of all transactions. For all uses it can be imperative that the advertised privacy guarantees are maintained. Consider an attacker whose goal is to determine whether Alice made a sensitive payment to a merchant. The attacker might have access to Alice's records at a currency exchange (e.g., through hacks or collusion), and/or to the records of the merchant. In this scenario, the pay-

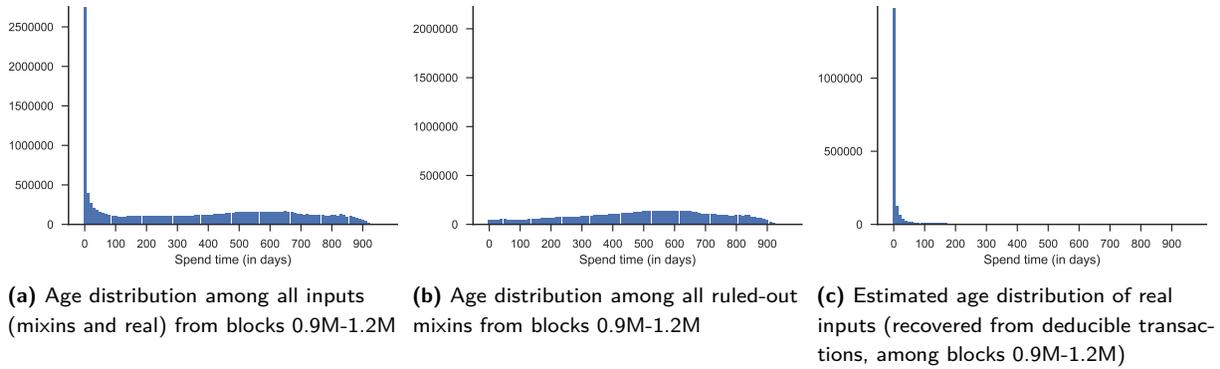

**(a)** Age distribution among all inputs (mixins and real) from blocks 0.9M-1.2M

**(b)** Age distribution among all ruled-out mixins from blocks 0.9M-1.2M

**(c)** Estimated age distribution of real inputs (recovered from deducible transactions, among blocks 0.9M-1.2M)

**Fig. 2.** Age distributions of Monero mixins. In each graph, the Y-axis is the number of TXOs, and the X-axis represents the time difference between input and referenced output in days. The left graph (a) shows the distribution of all transaction inputs from blocks 0.9M to 1.2M where at least 1000 possible TXOs are available to choose from. Graph (b) shows the age distribution among mixins that can be ruled out. Graph (c) shows that real TXOs (based on deducible transactions, see Section 3) typically conform to a highly-skewed distribution. The disparity between these distributions makes it possible to guess the real input with an estimated 80% accuracy.

ment system must prevent the attacker from confirming that money flowed from Alice to the merchant.

In this paper we conduct an empirical analysis of the Monero blockchain dataset. In particular, we evaluate the impact of two weaknesses in Monero's mixin sampling strategy, which substantially undermine its privacy guarantees. Even though neither of these weaknesses is entirely new, having been discussed by developers since as early as 2015, we find that users who made privacy-sensitive transactions prior to February 2017 are at significant risk of post hoc deanonymization. We estimate that more than 200,000 transactions from July 2016 to February 2017 may be affected.

### Weakness 1. Most Monero transaction inputs prior to February 2017 contain deducible mixins, and can be traced to prior transactions via analysis

The Monero software allows users to configure the default number of mixins to include in each transaction. Most Monero transaction inputs (64.04% of all transaction inputs) do not contain any mixins at all ("0-mixin transactions"), but instead explicitly identify the prior transaction output (TXO) they spend, much like ordinary Bitcoin transactions.

0-mixin transactions not only provide no privacy to the users that created them, but also present a privacy hazard if other users include the provably-spent outputs as mixins in other transactions. When the Monero client chooses mixins, it does not take into account whether the potential mixins have already been spent. We find that among Monero transaction inputs with one or more mixins, 63% of these are deducible, i.e. we can irrefutably identify the prior TXO that they spend.

### Weakness 2. Mixins are sampled from a distribution that does not resemble real spending behavior, and thus the real inputs can usually be identified

When the Monero client spends a coin, it samples mixins to include by choosing randomly from a triangular distribution over the ordered set of available TXOs with the same denomination as the coin being spent. However, when users spend coins, the coins they spend are not chosen randomly from the blockchain, but instead appear (based on our empirical observations) as though drawn from a highly skewed distribution.

In Figure 2 we show data from the Monero blockchain that confirms these assumptions. Figure 2(c) shows the real age of inputs in a representative subset of Monero transactions for which the real input is known (i.e., among deducible transaction inputs as described above). Figure 2(b) shows the age distribution of mixins for which we know that they are not real inputs. When looking at the overall distribution of all inputs, shown in Figure 2(a), the overall distribution can clearly be seen as a mixture of these two distributions. Among transactions for which we have ground truth (i.e., the deducible transaction shown in (c)), we find that *the real input is usually the "newest" input, 92.33% of the time*; based on simulation (Section 4), we estimate this holds for 80% of all transactions. Our results are summarized in Table 1.

**Table 1.** Traceability of Monero transaction inputs with 1+ mixins (up to block 1288774, excluding RingCT inputs). Deducible inputs can be traced with complete certainty to the transaction output they spend (see Section 3). Among deducible transaction inputs, the real input is usually the "newest" one (see Section 4). Entries marked [Est.] are estimated by extrapolating from deducible transaction inputs, under the assumption that the spend-time distribution of deducible transactions is representative of the distribution overall.

|  | **Not deducible** | **Deducible** | **Total** |
|---|---|---|---|
| Real input is not newest | 14.82% [Est.] | 302078 (4.83%) | 19.65% [Est.] |
| Real input is newest | 22.24% [Est.] | 3635253 (58.11%) | 80.35% [Est.] |
| Total | 2318273 (37.06%) | 3937331 (62.94%) | 6255604 (100%) |

## Monero usage

We turn to the ecosystem and analyze the behavior of intermediaries to better understand the transactions affected by these weaknesses. While the early days of Monero were dominated by mining activity, starting in 2016 we see substantial growth in transaction volume. After accounting for the estimated impact of mining pools, which opt-out of privacy by publishing their transactions on webpages, there remain a substantial number of potentially privacy-sensitive transactions, more than a thousand per day. We estimate that many of these transactions (about 25% of daily transactions at its peak) relate to the former underground marketplace AlphaBay, which positioned Monero as a more secure alternative to Bitcoin. The recent seizure of AlphaBay serves as a reminder of the fragility of these marketplaces, whether due to lawful actions, hacks, or exit scams [28], leaving users remain at risk of deanonymization. In Section 7 we discuss our results in the context of three high-profile criminal uses of Monero. Of course, Monero and other privacy-preserving technologies have legitimate uses as well, such as providing privacy for activists and supporting free speech within oppressive regimes.

In all of these, we consider an analyst model who has access to not just public blockchain data, but also records from exchanges and merchants, obtained through seizure or subpoena.

## Threat model

Alice purchases a quantity of XMR currency on a popular cryptocurrency exchange, such as Kraken or Poloniex, and withdraws it to a Monero wallet on her home computer, as recommended by Monero best practices guides [2]. Over time, she uses it for a number of innocuous activities such as online shopping and sending money to friends. In other words, she transacts with several parties using Monero under her real identity (the exchange because it is mandatory, the shopping site so as to have goods shipped to her, and so on). Later, Alice uses her wallet pseudonymously for a sensitive payment, one where she expects privacy. The attacker's goal is to link the pseudonymous account to a real name.

We consider a powerful attacker who is able to obtain two sets of logs (whether through hacks, seizures, or collusion). The first set of logs is of withdrawal transactions from the exchange (or another transaction where Alice used her real identity). The second set of logs is of deposits at the merchant where Alice made her sensitive payment. Due to the use of one-time addresses in Monero, the withdrawal transaction is the only piece of information on address ownership the attacker possesses. They cannot further infer common ownership of addresses (a deanonymization attack possible in many other cryptocurrencies [18, 24]). Furthermore, the attacker does not possess Alice's private keys and cannot break any cryptographic primitives.

The success of the attack depends on the attacker's ability to link the withdrawal transaction to one of the deposits at the merchant by deducing the real spend among Monero's chaff inputs, thereby confirming Alice as the originator of the payment. We note that in addition to our attack, an attacker who is able to use side channels such as additional timing information (e.g., a time zone) can further reduce the effective anonymity set provided by the mixins.

## Proposed countermeasures

We propose an improved mixin sampling strategy that can mitigate these weaknesses for future transactions. Our solution is based on sampling mixins according to a model derived from the blockchain data. We provide evidence that the "spend-time" distribution of real transaction inputs is robust (i.e., changes little over time and across different software versions), and can be well approximated with a simple two-parameter model. We extend the improved sampling with a sampling procedure which samples mixins in "bins." We show this binned sampling ensures privacy even in spite of a compromised sampling distribution.

**Lessons**

Our work highlights the special difficulties of designing privacy-preserving blockchain systems. As noted by Goldfeder et al. [12], cryptocurrency privacy combines the challenges of both anonymous communication and data anonymization. The data are necessarily public forever and vulnerabilities discovered at any point can be exploited to compromise privacy retroactively, as exemplified by the deducible Monero transactions from mid 2016. Privacy is also weakened by the fact that choices made by some users may affect other users detrimentally, such as the mining pools' practice of publishing payout transactions.

**Concurrent and related work on Monero traceability**

Two reports from Monero Research Labs (MRL-0001 [23] and MRL-0004 [16]) have previously discussed concerns about Weakness 1, known as the "chain-reaction" attack. However, MRL-0001 considered only an active attacker that must own coins used in previous transactions. Our results show this vulnerability is not hypothetical and does not require an active attack, but in fact leads to the traceability of most existing transactions to prior to February 2017. MRL-0004 [16] discussed a passive attack scenario and provided a simulation analysis predicting that the mandatory 2-mixin minimum (implemented in version 0.9) would "allow the system to recover from a passive attack quite quickly." Our results (Figure 5) show that indeed the fraction of deducible inputs indeed drops steadily after instituting the 2-mixin minimum (from 95% in March 2016 down to 20% in January 2017), though a significant fraction remain vulnerable.

Concurrently and inpendently of our work, Kumar et al. [15] also evaluated the deducibility attack in Monero. Our work differs from (and goes beyond) their analysis in four main ways. First, we account for correlation between the deducibility attack and temporal analysis in our simulation results (Section 4.3). Though deducibility has been addressed in current versions, temporal analysis remains independently effective. Second, our analysis of the Monero ecosystem (Section 5) refutes a possible objection that our analysis only applies to irrelevant transactions that do not need privacy (such as mining pool payouts). Third, while we propose a similar countermeasure to temporal analysis, we evaluate ours through simulation. Finally, our binned mixins countermeasure is novel and defends against a strong adversary with prior information.

# 2 Background

Since the inception of Bitcoin in 2009 [21], a broad ecosystem of cryptocurrencies has emerged. A cryptocurrency is a peer-to-peer network that keeps track of a shared append-only data structure, called a blockchain, which represents a ledger of user account balances (i.e., mappings between quantities of currency and public keys held by their current owner). To spend a portion of cryptocurrency, users broadcast digitally-signed messages called transactions, which are then validated and appended to the blockchain.

In slightly more detail, each cryptocurrency transaction contains some number of inputs and outputs; inputs consume coins, and outputs create new coins, conserving the total balance. Each input spends an unspent transaction output (TXO) created in a prior transaction. Together, these form a transaction graph.

The public nature of blockchain data poses a potential privacy hazard to users. Since each transaction is publicly broadcast and widely replicated, any potentially-identifying information can be data-mined for even years after a transaction is committed. Several prior works have developed and evaluated techniques for transaction graph analysis in Bitcoin [18, 24, 25]. Our present work shows that the Monero blockchain also contains a significant amount of traceable data.

The function of the peer-to-peer network and consensus mechanism is not relevant to our current work, which focuses only on blockchain analysis; readers can find a comprehensive overview in Bonneau et al. [5]. Network-based forensic attacks are also known to threaten privacy in Bitcoin [3, 14], but applying this to Monero is left for future work.

**Cryptonote: Non-interactive mixing with ring signatures**

The Cryptonote protocol [30] introduces a technique for users to obscure their transaction graph, in principle preventing transaction traceability. Instead of explicitly identifying the TXO being spent, a Cryptonote transaction input identifies a set of possible TXOs, including both the real TXO along with several chaff TXOs, called mixins (as illustrated in Figure 1). Instead of an ordinary digital signature, each Cryptonote transaction comes with a ring signature (a form of zero-knowledge proof) that is valid for one of the indicated TXOs, but that does not reveal any information about which one is real. To prevent double-spending, every input must provide a key image that is unique to the output being spent, and the network must check whether this key image has ever been revealed before.

Several cryptocurrencies are based on the Cryptonote protocol, including Monero, Boolberry, Dashcoin, Bytecoin, and more. We focus our empirical analysis on Monero, since it is currently the largest and most popular, e.g. it has the 12th largest market cap of all cryptocurrencies, over \$750M. However, our results are also applicable to other Cryptonote-based protocols (as we show for Bytecoin in Appendix C).

### Choosing mixin values in Cryptonote

The Cryptonote protocol does not provide an explicit recommendation on how the "mixins" should be chosen. However, the original Cryptonote reference implementation included a "uniform" selection policy, which has been adopted (at least initially) by most implementations, including Monero. Since all the TXOs referenced in a transaction input must have the same denomination (i.e., a 0.01 XMR input can only refer to an 0.01 XMR output), the client software maintains a database of available TXOs, indexed by denomination. Mixins are sampled from this ordered list of available TXOs, disregarding any temporal information except for their relative order in the blockchain.

In principle, it is up to an individual user to decide on a policy for how to choose the mixins that are included in a transaction. Since it is not a "consensus rule," meaning that miners do not validate that any particular distribution is used, clients can individually tune their policies while using the same blockchain. The Monero command-line interface allows users to specify the number of mixins, with a current default of 4.

Over the past several years, Monero's mixin selection policy has undergone several changes; we describe the important ones below. A summary of the timeline relevant to our data analysis is shown in Figure 3.

**Prior to version 0.9.0 (January 1, 2016)** In the initial Monero implementation, mixins were selected uniformly from the set of all prior TXOs having the same denomination as the coin being spent. As a consequence, earlier outputs were chosen more often than newer ones.

**After version 0.9.0 (January 1, 2016)** Version 0.9.0 introduced a new policy for selecting mixins based on a triangular distribution, favoring newer coins as mixins over older ones. This change was motivated by the belief that newer inputs are more likely to be the real input [9]. This version also introduced a mandatory minimum number of 2 mixins per transaction input, as recommended by MRL-0001 [23]. This mandatory minimum was enforced after a "hard fork" flag day, which occurred on March 23, 2016.

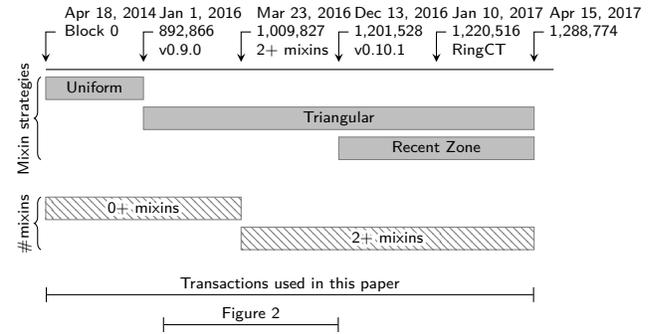

**Fig. 3.** Data considered in our experiment.

**After version 0.10.0 (September 19, 2016)** Version 0.10.0 introduced a new RingCT feature [22], which allows users to conceal the denomination of their coins, avoiding the need to partition the available coins into different denominations and preventing value-based inference attacks. RingCT transactions were not considered valid until after a hard fork on January 10, 2017.

The RingCT feature does not directly address the traceability concern. But as RingCT transactions can only include other RingCT transaction outputs as mixins, and since it was deployed after the 2-mixin minimum took effect (in version 0.9.0), there are no 0-mixin RingCT inputs to cause a hazard.

**After version 0.10.1 (December 13, 2016)** Version 0.10.1 included a change to the mixin selection policy: now, some mixins are chosen from among the "recent" TXOs (i.e., those created within the last 5 days, called the "recent zone"). Roughly, the policy is to ensure 25 % of the inputs in a transaction are sampled from the recent zone.

**After version 0.11.0 (September 07, 2017)** Version 0.11.0 increased the minimum number of mixins per transaction input to 4, which was enforced after a hard fork on September 15, 2017. This version also incorporates temporal analysis countermeasures (increasing the number of mixins chosen from the recent zone, and narrowing the recent zone from 5 days to 3 days) based on recommendations from an early draft of our paper.

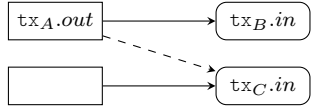

**Fig. 4.** 0-mixins effectively reduce the untraceability of other transactions: the dashed reference can be ruled out since $\mathtt{tx}_A.out$ must have been spent in $\mathtt{tx}_B.in$.

### Transaction notation

We briefly introduce some notation for describing transaction graphs. For a transaction $\mathtt{tx}$, $\mathtt{tx}.in$ denotes the vector of $\mathtt{tx}$'s transaction inputs, and $\mathtt{tx}.out$ denotes the vector of $\mathtt{tx}$'s transaction outputs. We use subscripts to indicate the elements of input/output vectors, e.g. $\mathtt{tx}.in_1$ denotes the first input of $\mathtt{tx}$. Each Cryptonote transaction input contains a reference to one or more prior transaction outputs. We use array notation to denote the individual references of an input. We use a dashed arrow, $\dashrightarrow$, to denote this relationship, e.g. $\mathtt{tx}_A.out_i \dashrightarrow \mathtt{tx}_B.in_j[m]$ means that the $m$'th reference of the $j$'th input of transaction $\mathtt{tx}_B$ is a reference to the $i$'th output of $\mathtt{tx}_A$. Although a Cryptonote transaction input may contain more than one reference, only one input is the *real* reference (known only to the sender), indicated by a solid arrow. Thus $\mathtt{tx}_A \to \mathtt{tx}_B$ indicates that $\mathtt{tx}_A$ contains an output that is spent by one of the inputs in $\mathtt{tx}_B$.

Transactions included in the blockchain are processed in sequential order; we use $\mathtt{tx}_A < \mathtt{tx}_B$ to indicate that $\mathtt{tx}_A$ occurs before $\mathtt{tx}_B$. Other properties of a transaction are defined as functions, and introduced as needed. For example, $\mathsf{time}(\mathtt{tx})$ refers to the timestamp of the block in which $\mathtt{tx}$ is committed.

## 3 Deducible Monero Transactions

A significant number of Monero transactions do not contain any mixins at all, but instead explicitly identify the real TXO being spent. Critically, at the beginning of Monero's history, users were allowed to create zero-mixin transactions that do not contain any mixins at all. Figure 5 shows the fraction of transactions containing zero-mixin inputs over time. As of April 15, 2017 (at block height 1288774), a total of 12158814 transaction inputs do not contain any mixins, accounting for 64.04% of all inputs overall.

One might think at first that 0-mixin transactions are benign. Transactions with fewer mixins are smaller, and hence cost less in fees; they thus represent an eco-nomical choice for an individual who does not explicitly desire privacy for a particular transaction. However, it turns out that the presence of 0-mixin transactions is a hazard that reduces the untraceability of other transactions, even those that include one or more mixins. For example, as shown in Figure 4, suppose a transaction output $\mathtt{tx}_A.out_i$ is spent by a 0-mixin transaction input $\mathtt{tx}_B.in_i$ (i.e. $\mathtt{tx}_A.out_i \dashrightarrow \mathtt{tx}_B.in_j$ where $|\mathtt{tx}_B.in_j| = 1$, from which we can conclude $\mathtt{tx}_A.out_i \to \mathtt{tx}_B.in_j$). Now, suppose $\mathtt{tx}_A.out_i$ is also included as a mixin in a second transaction with one mixin, $\mathtt{tx}_A.out_i \dashrightarrow \mathtt{tx}_C.in_k[m]$, where $|\mathtt{tx}_C.in_k| = 2$. Since we know that the given output was actually spent in $\mathtt{tx}_B$, we can deduce it is not spent by $\mathtt{tx}_C$ (i.e., $\mathtt{tx}_A.out_i \not\to \mathtt{tx}_C$), and hence the remaining input of $\mathtt{tx}_C.in_j$ is the real one.

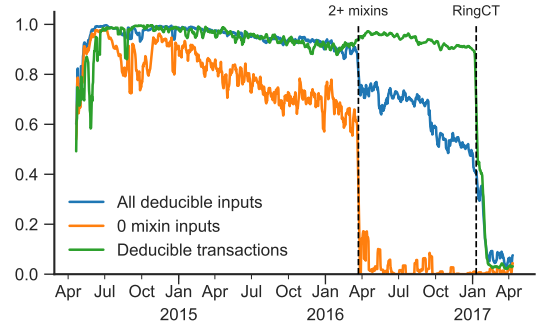

**Fig. 5.** Fraction of transaction inputs that can be deduced and transactions including at least one deducible input (averaged over intervals of 7 days)

Notice that in this example, it does not matter if the real spend $\mathtt{tx}_B$ that renders it deducible occurs after the 1-mixin transaction $\mathtt{tx}_C$, i.e., if $\mathtt{tx}_C < \mathtt{tx}_B$. Thus at the time $\mathtt{tx}_C$ is created, it is impossible to know whether a future transaction will render that mixin useless. However, the problem has been exacerbated by the behavior of the Monero client software, which does not keep track of whether a potential mixin has already been clearly spent, and naïvely includes degenerate mixins anyway.

### 3.1 Implementation

We extracted relevant information from the Monero blockchain, up to block 1288774 (April 15, 2017) and stored it in a Neo4j graph database (11.5GB of data in total). We then apply the insight above to build an iterative algorithm, where in each iteration we mark all of the mixin references that cannot be the real spend since we have already deduced that the corresponding output has been spent in a different transaction. With each iteration, we further deduce the real inputs among additional

**Table 2.** Monero transaction inputs where the real input can be deduced (1+ mixins, $\geq 1000$ TXOs available, excluding RingCT).

| | Before 2-mixin hardfork | | | After 2-mixin hardfork | | | After 0.10.1, prior to Apr 15, 2017 | | |
|---|---|---|---|---|---|---|---|---|---|
| | Total | Deducible | (%) | Total | Deducible | (%) | Total | Deducible | (%) |
| 1 mixins | 683458 | 608087 | (88.97) | 0 | – | – | 0 | – | – |
| 2 mixins | 250520 | 206276 | (82.34) | 1882681 | 1209259 | (64.23) | 732251 | 308926 | (42.19) |
| 3 mixins | 634520 | 480500 | (75.73) | 564525 | 376920 | (66.77) | 126795 | 65738 | (51.85) |
| 4 mixins | 217493 | 156767 | (72.08) | 376432 | 192348 | (51.10) | 145687 | 33022 | (22.67) |
| 5 mixins | 87077 | 43214 | (49.63) | 48806 | 26599 | (54.50) | 3900 | 950 | (24.36) |
| 6 mixins | 115199 | 65546 | (56.90) | 224202 | 119716 | (53.40) | 24817 | 7890 | (31.79) |
| 7 mixins | 3671 | 1680 | (45.76) | 4499 | 1770 | (39.34) | 1711 | 235 | (13.73) |
| 8 mixins | 2216 | 1067 | (48.15) | 5048 | 1968 | (38.99) | 1458 | 249 | (17.08) |
| 9 mixins | 1811 | 838 | (46.27) | 3264 | 1069 | (32.75) | 264 | 48 | (18.18) |
| 10+ mixins | 57363 | 11997 | (20.91) | 46791 | 12970 | (27.72) | 9145 | 1682 | (18.39) |
| Total Overall | 2053328 | 1575972 | (76.75) | 3156248 | 1942619 | (61.55) (62.94) | 1046028 | 418740 | (40.03) |

transaction input sets. (An alternative formulation – as a SAT problem – is presented in Appendix A.)

## 3.2 Results on Deducible Transactions

Table 2 shows the results from applying the approach described above to Monero blockchain data. As it turns out, approximately 63% of Monero transaction inputs (with 1+ mixins) so far can be traced in this way.

In Figure 6, we show how the vulnerability of Monero transactions to deduction analysis varies with the number of mixins chosen, and in Figure 5 we show how this has evolved over time. We see that transactions with more mixins are significantly less likely to be deducible, as one would hope. Even among transactions with the same number of mixins, transactions made with later versions of the software are less vulnerable. This is because at later dates, especially after the network started to enforce a minimum of 2 mixins, the 0-mixin transaction outputs accounted for a smaller number of the available mixins to choose from. However, the share of transactions with at least one deducible input (enough for an attacker to link an account to a user) stays above 80% even after the 2-mixins minimum. Surprisingly, we found over 100,000 transaction inputs with 10 or more mixins, presumably indicating a high level of desired privacy, that are vulnerable under this analysis.

**Applicability to current and future transactions using RingCT**
The weakness studied in this section is primarily a concern for transactions made in the past, as transactions using the new RingCT transaction option are generally immune. RingCT has been available to users since January 2017, and at the time of writing has already been widely deployed. RingCT transactions are now used by default for the vast majority of new transactions. The reason why these new transactions are immune is not because of the RingCT mechanism itself, but rather because RingCT was only deployed after the mandatory 2-mixin minimum was enforced. Therefore, RingCT transaction outputs cannot be spent by 0-mixin inputs.

**Applicability to other Cryptonote cryptocurrencies**
As the deducibility attack originates from the mixing sampling procedure inherent to the Cryptonote protocol, other cryptocurrencies based on Cryptonote share the same weaknesses. Appendix C contains results of an analysis of Bytecoin, an early implementation of the Cryptonote protocol. We deduce the real input for 29% of transaction inputs with 1 or more mixins. While this rate is smaller than in Monero, we conjecture it to be a result of lower usage of Bytecoin.

## 4 Tracing With Temporal Analysis

In the previous section, we showed that a majority of Monero transactions inputs can be traced with certainty through logical deduction. In this section, we investigate an entirely unrelated complementary weakness which traces inputs probabilistically.

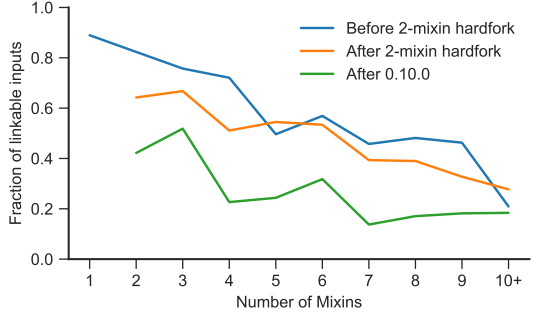

**Fig. 6.** Transaction inputs are less likely to be deducible if they have more mixins and if they are found among later blocks in the Monero blockchain.

## 4.1 Effective-Untraceability

To quantify the untraceability of a transaction input when some referenced outputs are more likely to be the real spend we first define the effective-untraceability set size or the "effective-untraceability."

Anonymity set size is a standard privacy metric which typically assumes each element in the anonymity set is equiprobable. This is not the case in Monero as the temporal analysis attacks we will demonstrate show that some referenced outputs are more likely than others to be the real spend. Following the approach of [7, 26] we use entropy to account for these differing probabilities. We measure guessing entropy rather than Shannon entropy [27] because it is intuitive to our setting, an attacker trying to guess the real spend, and is easily relatable to effective-untraceability.

First defined in [17], guessing entropy is commonly used as a measure of password strength [4]. In the context of untraceability guessing entropy is the expected number of guesses before guessing the spent output. A transaction input's guessing entropy is

$$\text{Ge} = \sum_{0 \leq i \leq M} i \cdot p_i$$

where $p = p_0, p_1, \ldots, p_M$ are probabilities, sorted highest to lowest, that a referenced output is the real spend of a transaction input.

We define the effective-untraceability, i.e. effective-anonymity set size, of a transaction input as $(1+2\cdot\text{Ge})$. If all referenced outputs of a transaction input are equally likely to be the real spend, the effective-untraceability for that input is $M + 1$.

## 4.2 The Guess-Newest Heuristic

Among all the prior outputs referenced by a Monero transaction input, the real one is usually the newest one.

Figure 2(c) shows the spend-time distribution for deducible transaction inputs (i.e., zero-mixin inputs and inputs for which the real TXO can be deduced using the technique from Section 3). As can be seen, this distribution is highly right-skewed; in general users spend coins soon after receiving them (e.g., a user might withdraw a coin from an exchange, and then spend it an hour later). In contrast, the distribution from which (most) mixins are sampled (either a uniform distribution or a triangular distribution, for the most part) includes much older coins with much greater probability.

**Cross-validation with deducible transactions**
In order to quantify the effect of these mismatched distributions, we examine the rank order of the transactions with 1 or more mixins for which we have ground truth (i.e. the deducible transactions from Section 3). Table 3 shows the percentages of deducible transaction inputs for which the real (deduced) reference is also the "newest" reference. It turns out that overall, 92% of the deducible inputs could also be guessed correctly this way. For $1 \leq M \leq 10$ such transaction inputs have an effective-untraceability of no more than 1.16–1.80.

We note that transactions with more mixins are only slightly less vulnerable to this analysis: even among transaction inputs with 4 mixins (the required minimum since September 2017) the Guess-Newest heuristic still applies to 81% of these transactions.

**Validation with ground truth**
We also verify the heuristic using our own ground truth, which we obtain by periodically creating transactions using the default wallet in Monero 0.10.3.1. We set up wallets for three scenarios and send transactions from one wallet to another. The time gap between two transactions follows an exponential distribution, with the rate parameter set such that the means of the distributions correspond to 30 minutes, 4 hours, and 1 day.

We evaluate whether guessing the most-recently created input identifies the true input. At a 95%-confidence level, we get success probabilities of $0.95 \pm 0.02$ for the 30-minute interval ($n = 120$), $0.90 \pm 0.03$ for 4 hours ($n = 84$), and $0.42 \pm 0.14$ for 24 hours ($n = 12$).

**Table 3.** Percentage of deducible transaction inputs where the real input is the "newest" input.

| | Before 2-mixin hardfork | | | After 2-mixin hardfork | | | After 0.10.1, prior to Apr 14, 2017 | | |
|---|---|---|---|---|---|---|---|---|---|
| | Deducible | Newest | (%) | Deducible | Newest | (%) | Deducible | Newest | (%) |
| 1 mixins | 608087 | 585424 | (96.27) | 0 | – | – | 0 | – | – |
| 2 mixins | 206276 | 191372 | (92.77) | 1209259 | 1126924 | (93.19) | 308926 | 293051 | (94.86) |
| 3 mixins | 480500 | 461154 | (95.97) | 376920 | 353246 | (93.72) | 65738 | 59693 | (90.80) |
| 4 mixins | 156767 | 139626 | (89.07) | 192348 | 149722 | (77.84) | 33022 | 18889 | (57.20) |
| 5 mixins | 43214 | 39854 | (92.22) | 26599 | 24971 | (93.88) | 950 | 473 | (49.79) |
| 6 mixins | 65546 | 51816 | (79.05) | 119716 | 102378 | (85.52) | 7890 | 6458 | (81.85) |
| 7 mixins | 1680 | 1522 | (90.60) | 1770 | 989 | (55.88) | 235 | 115 | (48.94) |
| 8 mixins | 1067 | 964 | (90.35) | 1968 | 1310 | (66.57) | 249 | 163 | (65.46) |
| 9 mixins | 838 | 692 | (82.58) | 1069 | 355 | (33.21) | 48 | 40 | (83.33) |
| 10+ mixins | 11997 | 10822 | (90.21) | 12970 | 11750 | (90.59) | 1682 | 1480 | (87.99) |
| Total | 1575972 | 1483246 | (94.12) | 1942619 | 1771645 | (91.20) | 418740 | 380362 | (90.83) |
| Overall | | | | | (92.33) | | | | |

## 4.3 Monte Carlo Simulation

The Guess-Newest heuristic and the deducibility attack (Section 3) are not entirely independent. The transactions that are deducible tend to be those that include old mixins, which thus also makes them more likely susceptible to Guess-Newest. As a consequence, cross-validation of the heuristic using only deducible transactions overstates the effectiveness of temporal analysis.

To control for this correlation, we also employ an alternative validation strategy based on a Monte Carlo simulation. In each trial, we simulate a transaction, where the real input is sampled from the dataset of deducible transactions, but the mixins are chosen using the sampling algorithms in the Monero software. Two factors determine which mixins are available to choose from: the denomination of the real input, and the blockchain data at the time the transaction is created. To simplify the simulation, we fix the block height at 1236196 (Jan 31, 2017), We consider the dataset of deducible and 0-mixin transactions as records of the form ($\mathsf{spendtime}, \mathsf{denomination}$), where the spend time is the block timestamp difference between when a transaction output is created and when it is spent, i.e. $\mathsf{spendtime}(\mathtt{tx}_B.in_j) = \mathsf{time}(\mathtt{tx}_B) - \mathsf{time}(\mathtt{tx}_A)$, where $\mathtt{tx}_A \to \mathtt{tx}_B$. We sample uniformly from these records to choose the real input. Next we apply the mixin sampling strategy using the available transaction outputs matching the chosen denomination. In Appendix B, we provide the pseudocode for the sampling procedures used in our simulation. We note that our sampling procedures are simplified models, and in particular elide the handling of edge cases such as avoiding "locked" coins that have been recently mined.

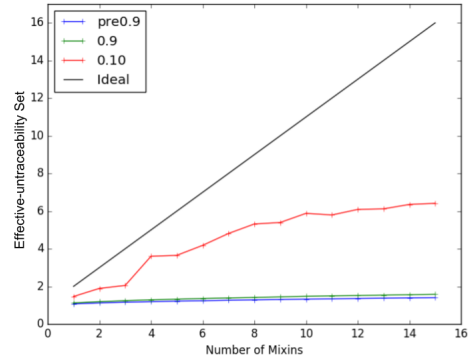

**Fig. 7.** Estimated vulnerability to the Guess-Newest heuristic for varying sampling policies in Monero, based on Monte Carlo simulations (100k trials each).

In Figure 7 we show the results of simulating transactions using the sampling rules from the three main Monero versions (pre 0.9, 0.9, and 0.10.1), as applied to the blockchain at height 1236196 (Jan 31, 2017). For versions 0.9.0 and prior, even up to 4 mixins, our simulation suggests that the newest input can be guessed correctly 75% of the time. Across all versions, including more mixins does reduce the effectiveness of this heuristic, although the benefit of each additional mixin is significantly less than the ideal.

Each subsequent update to Monero's mixin sampling procedure has improved the resistance of transactions to the Guess-Newest heuristic. We note that from developer discussions [10] it appears that this concern has indeed been the motivation for such changes. In particular, the triangular distribution was adopted in place of the uniform distribution in Monero version 0.9 specifically because it chooses new mixins with higher probability than "old" mixins, and version 0.10.1 explicitly introduces a policy that often includes additional "recent" (within 5 days) mixins. However, we believe the magnitude of the problem has been underestimated. Under the current default behavior, i.e. 4 mixins, and using the 0.10.1 sampling procedure, we estimate that the correct input reference can be guessed with 45% probability (rather than the 20% ideal if all input references were equally likely).

Excluding all of the deducible inputs, the average number of mixins among the remaining inputs is 3.53. We therefore estimate that as a lower bound, based on the "0.10.1" line in Figure 7, that at least 60% of the remaining inputs are correctly traced using Guess-Newest. Extrapolating from this figure, we estimate that in total the Guess-Newest heuristic correctly identifies 80% of all Monero inputs (Table 1).

## 5 Characterizing Monero Usage

In the previous section, we showed that a significant number of Monero transactions are vulnerable to tracing analysis. However, not all transactions are equally sensitive to privacy. We consider typical or *normal* transactions in Monero to be privacy-sensitive. On the other hand, *public* transactions, for which privacy is not a major concern to their participants, and where details of the transaction may even be publicly disclosed, form a special case. We next quantify these different usage types for Monero transactions, to assess potential impact of our attacks on privacy-conscious users.

### 5.1 Quantifying Mining Activity

An integral part of a cryptocurrency is mining, which refers to the process of bundling transactions in blocks and minting new currency. Whenever a block is created, the "miner" of that block receives a monetary reward, described by a special "coinbase" transaction. With an initial block interval of 1 minute until March 20, 2016, and a block interval of 2 minutes thereafter, it is conceivable that – at least in the early days – a significant share of daily transactions relate to mining.

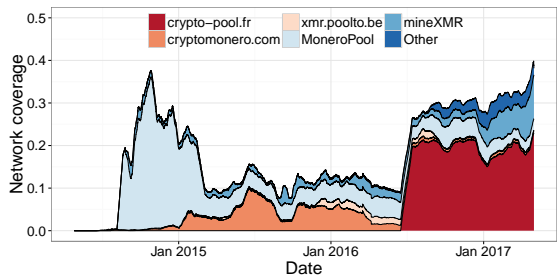

**Fig. 8.** Fraction of blocks created by mining pools for which payment transactions are made public and have been collected.

Miners often combine their efforts by joining mining pools to reduce the variance of their payouts. Usually, the pool owner receives the full block reward and then distributes the reward according to each miner's contribution. To provide transparency and accountability, many pools publicly announce the blocks they find as well as the payout transactions in which they distribute the rewards. This, however, reveals sensitive information about the relation between payout and coinbase transactions. If an input in a pool's payout transaction spends from a coinbase transaction of a block known to belong to the same pool, it is likely the real spend. This deducibility may even weaken the privacy of other transactions, similar to a 0-mixin input. Transactions related to mining activity should thus be considered public.

To account for public transactions related to mining, we crawled the websites of 18 Monero mining pools and extracted information about 73,667 pool payouts. We also collected information about 210,800 non-orphaned blocks that had been won by these pools. We analyzed the coverage of this mining activity by computing the fraction of all blocks in the network that had been produced by pools in our dataset over a moving window of 20,000 blocks. Figure 8 shows the results of this analysis.

Our data accounts for roughly 30% of the Monero mining power in late 2014. In early 2015, according to an archive of Monerohash.com [20] retrieved via The Wayback Machine [29], over 70% of the mining power belonged to a combination of the Dwarf [8] and MinerGate [19] mining pools, neither of which reveals payment transactions, and of unknown mining sources. This centralization of hash power continued until April of 2016 at which point Dwarf and MinerGate were still significant (about 35%), but there was no longer significant unaccounted for mining power.

Besides the 73,667 transactions for which we have ground truth, we can estimate the number of unlabeled transactions in the network that are used for mining payments. To minimize transaction fees, most pools will pay their miners only a few times per day, batching together payments to many miners into a few transactions with up to hundreds of outputs. Most pools do offer an option for immediate payout upon request, but charge a significant fee for doing so. Additionally, pools will allow users to be paid to an exchange service instead

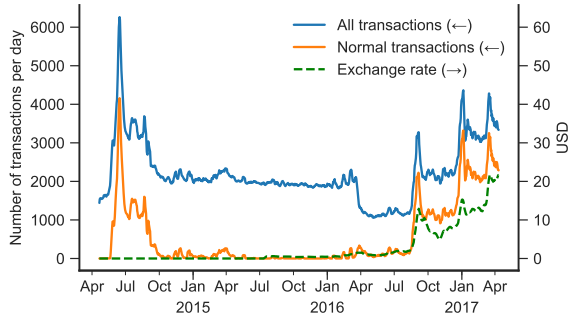

**Fig. 9.** Comparison of the volume of estimated non-mining transactions to the overall transaction volume.

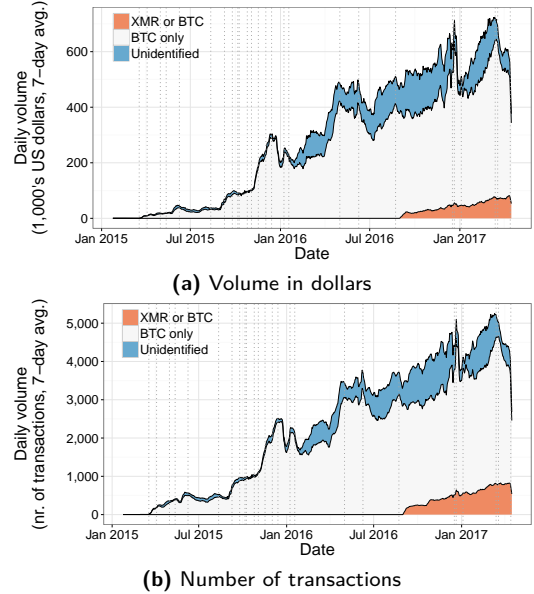

**(a)** Volume in dollars

**(b)** Number of transactions

**Fig. 10.** Estimated AlphaBay sales volume since inception

of to their own wallet. Thus, determining precisely the number of mining transactions is challenging.

Working under the assumption that the distribution of payments from the pools we observed tracks that of the pools that we were not able to observe, namely that the distribution of transactions is driven primarily by the needs and demands of the miners, we can estimate the volume of total mining-related transactions.

To do so, we compute the ratio between the total number of transactions observed and the total number of blocks mined by pools in our dataset. We estimate that in addition to the coinbase transaction, approximately 0.438 transactions related to mining payments are created for each new block in Monero.

Figure 9 plots the number of overall transactions per day against the number of estimated normal transactions. Non-mining use was generally low during 2015 and until mid-2016, with at most a few hundred transactions per day. Transaction volume increases after mid-2016, rising over 2,000 per day in 2017, which cannot be accounted for by mining-related activity. There is also strong correlation between the number of normal transactions and the Monero exchange rate. This suggests that changes in the ecosystem might be responsible for the additional transactions. We next explore a potential source for these unaccounted for normal transactions.

## 5.2 Usage on Online Anonymous Marketplaces: the AlphaBay Case

Online anonymous marketplaces have traditionally used Bitcoin for monetary transactions [6], which makes them potentially vulnerable to transaction graph analysis [18]. Because Monero advertises stronger anonymity properties than Bitcoin, a couple of online anonymous marketplaces (Oasis, AlphaBay) have started supporting it [13].

AlphaBay, in particular, started its operations in December 2014 and featured tens of thousands of pages until it was shut down by an international law enforcement operation in July 2017, making it one of the longest-lived (and possibly largest) marketplaces to this date.

On August 22, 2016, AlphaBay announced that it would support Monero, and allowed vendors to list items accepting Monero. On September 1, 2016, buyers were then able to use Monero to purchase items on AlphaBay. Inspecting the number of daily transactions in Figure 9 reveals a strong correlation between these events and the overall transaction volume. On August 22, the day of the announcement, the number of transactions in the Monero blockchain increased by more than 80% compared to the previous day, and it peaked at 4,444 transactions on September 3, two days after Monero payments were made available on AlphaBay.

By default, an item listed on AlphaBay only accepts Bitcoin as a payment mechanism. Vendors have to explicitly allow Monero for it to be considered—and can also disable Bitcoin in the process. There are thus three types of items: items that only accept Monero, items that only accept Bitcoin, and items that accept both.

We obtained AlphaBay crawl data from Soska and Christin [28], and multiplied item feedback instances by item prices, to estimate sales volumes. We plot our results in Figure 10. Figure 10(a) shows the volume in US dollars, accounting for daily changes in Monero exchange rate. Each vertical dashed line corresponds to a scrape of the website. The top curve corresponds to the total daily volume of transactions, averaged over 7-day intervals. While sales volume remained fairly modest until mid-2015, it has steadily climbed to reach approximately USD 600,000 a day in 2017, which is more than the combined volume of the major online anonymous marketplaces in 2013–2015 [28].[1] Of those transactions, we are able to identify that a vast majority (in white) only accept Bitcoin. Starting in September 2016, a modest, but increasing number of items started accepting Monero along Bitcoin. The total amount of sales for these items, represented in orange, gives an upper bound for the dollar amount of Monero transactions on AlphaBay—as of early 2017, approximately USD 60,000/day (or 10% of all AlphaBay transactions in volume). To get a lower bound on the amount of Monero transactions, we also look at items that *only* accept Monero, but these remain negligible (totaling between 0 and USD 100/day in transactions). In short, up to USD 60,000/day of transactions on the AlphaBay marketplace used Monero.

Figure 10(b) plots the number of transactions over time (averaged, again, over 7-day rolling windows). We see that, as of early 2017, up to 1,000 transactions per day on AlphaBay might be relying on Monero. Comparing with Figure 9, we estimate that at most (approximately) a quarter of all Monero transactions could be accounted for by deposits at AlphaBay. Among the remaining 75% of Monero transactions that we estimate are non-mining, we cannot conclude how many are privacy sensitive. Good candidates for the sources of these transactions include, gambling, exchanges, and traditional direct user to user payments.

# 6 Countermeasures

We now propose two countermeasures to improve Monero's resistance to temporal analysis. The first is to improve the mixin-sampling procedure to better match the real spend-time of Monero users. The second takes a more pessimist view and aims to preserve some untraceability even in the face of a highly compromised mixin sampling distribution.

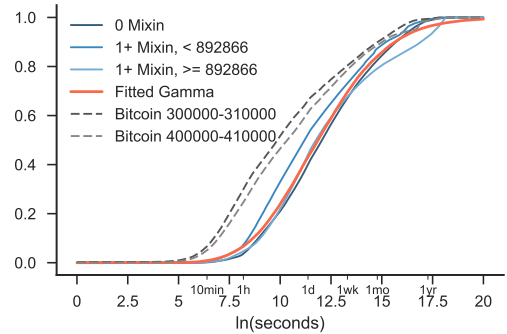

**Fig. 11.** CDFs of spend-time distributions in Bitcoin and in Monero (deducible transaction inputs), over multiple time intervals. A gamma distribution (red line) is fitted to the combined Monero data (shape=19.28, rate=1.61).

## 6.1 Fitted Mixin Sampling Distribution

In this section we discuss a way to improve the sampling procedure. At a high level, the idea is to estimate the actual spend-time distribution, and then sample mixins according to this distribution.

**Estimating the spend-time distribution**
In Figure 11, we show the CDFs of the spend-time distributions from the Monero blockchain as well as the Bitcoin blockchain. For Monero, we split the data by 0-mixin transaction inputs as well as the 1+ mixin inputs for which we can deduce the real input. From this graph, we make the following qualitative observations:

– Observation 1: The spend-time distribution appears invariant with respect to time.
– Observation 2: The spend-time distribution appears the same for 0-mixin as well as 1+ mixin transactions.
– Observation 3: While the Bitcoin spend-time distribution appears to have a somewhat similar shape, the distributions are quite different (i.e., the Kolmogorov-Smirnov distance is approximately 0.3).

Based on these observations, we set about fitting a parametric model to the combined Monero data. We heuristically determined that the spend time distributions, plotted on a log scale, closely match a gamma distribution. We used R's `fitdistr` function to fit a gamma distribution to the combined Monero data from deducible transaction inputs (in log seconds). The resulting best-fit distribution has shape parameter 19.28, and rate parameter 1.61. By inspection (Figure 11), this appears to accurately fit the overall Monero spend-time distribution.

---

**Sampling mixins using the spend-time distribution**

To make use of the spend-time distribution described above, we need a way of sampling transaction output indices that matches the ideal spend-time distribution as closely as possible. Our proposed method is to first sample a target timestamp directly from the distribution, and then find the nearest block containing at least one RingCT output. From this block, we sample uniformly among the transaction outputs in that block. This procedure is defined Algorithm 1.

To quantify the effectiveness of our new sampling scheme, we turn again to the Monte Carlo simulation described in Section 4.3. Figure 12 shows the effective untraceability set under several regimes, including the RingCT protocol as of 0.10.1 (the version prior to the initial preprint release of our paper), our proposed mixing sampling routine method (Ppd), and version 0.11.0 (which incorporates a countermeasure based on our preprint). Our proposed countermeasure performs very close to the ideal. At the default setting of 4 mixins, our method nearly doubles the effective untraceability set versus 0.10.1 (approximately $\approx 4$, 80% of the ideal, instead of approximately $\approx 2$); for large numbers of mixins, our improvement is nearly four times better. As the simulation is based on the same dataset to which we fit a parametric distribution, this is best understood as a goodness-of-fit test. Under this analysis, we also find that the countermeasure employed in 0.11.0 performs nearly as well as our proposed countermeasure at the default setting, though the gap increases with the ring size.

In the conference version of this paper, we evaluated our countermeasure using the results of several simulations based on an extrapolation that the transaction rate of Monero would remain constant. At the time of the experiment, March 2017, there were an average of 8.29 RingCT transaction outputs per block. We performed this extrapolation because unlike the deducibility weakness, which improves over time (see Figure 6), the problem of sampling from the wrong temporal distribution becomes *worse* over time, since the set of "old" mixins to choose from grows larger and larger. Our simulations with extrapolated data performed similarly to our experiments with current data from March 2018, although the effective untraceability set now is slightly better than projected.

---

**Algorithm 1.** Our proposed mixin sampling scheme.

SampleMixins(*RealOut, NumMixins*)

> MixinVector := [];
> **while** |MixinVector| < BaseReqMixCount **do**
> > $t \leftarrow$ Exp(GammaSample(*shape=19.28, rate=1.61*));
> > **if** $t >$ CurrentTime **then**
> > > continue;
> >
> > Let $B$ be the block containing at least one RingCT output, with timestamp closest to CurrentTime-$t$;
> > $i \leftarrow$ uniformly sampled output among the RingCT outputs in $B$ **if** $i \notin$ MixinVector and $i \neq$ RealOut.$idx$ **then**
> > > MixinVector.append($i$);
>
> **return** *sorted*(MixinVector + [RealOut.idx]);

---

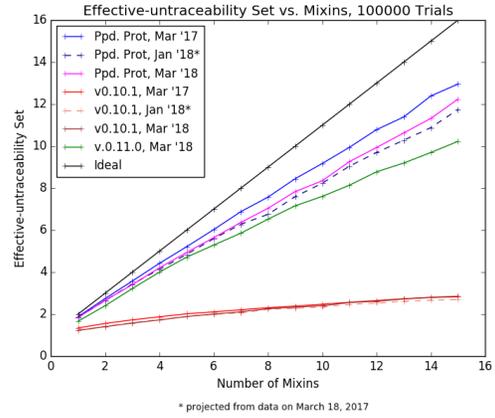

**Fig. 12.** Projection of the Guess-Newest vulnerability, and the improvement due to our proposed scheme.

## 6.2 Binned Mixin Sampling

Here we introduce a countermeasure called binned mixin sampling which modifies the current mixin sampling procedure. It is designed to maintain a minimum level of untraceability even in face of compromised mixin sampling distributions or deduction attacks. Binned mixin sampling improves on the current single mixin sampling, which offers no guarantee that temporal analysis will not completely trace a transaction input by reducing effective-untraceability to $\approx 1$.

**Compromised mixin sampling distributions**

Binned mixin sampling is designed to hedge against a mixin sampling distribution which poorly matches expected spend-time behavior. This is important because even if the mixin sampling distribution matches the overall spend-time distribution, the spend-time behavior of some groups may differ greatly. This difference, if known, could be leveraged to perform temporal analysis.

Recall our threat model (Section 1) where a forensic attacker Eve has withdrawal records of exchanges as well as deposit records at a merchant site (e.g., AlphaBay). Her goal is to trace Alice's withdrawal to a deposit at AlphaBay by confirming that it was used as an input of a deposit transaction. Unlike attacks discussed so far, Eve has additional information about Alice, Eve knows Alice's timezone, i.e. roughly the time of day when Alice receives and sends payments. Eve uses this information to determine the likelihood that each referenced output is the real spend, e.g., Ge = $(p_0 = 0.80, p_1 = 0.17, p_2 = 0.02, p_3 = 0.01) = 0.24$ results in Alice's output having an effective-untraceability of $2 \times (0.24) + 1 = 1.48$ (defined in Section 4.1). However, using binned mixin sampling the same number of mixins would ensure an effective-untraceability $\geq 2.0$.

**Binned mixin sampling**

As shown in Figure 13 our strategy is to group outputs in the Monero blockchain into sets of some fixed size, called bins, such that each output in a bin is confirmed in the same block or a neighboring block. This ensures each output in a bin has roughly the same age. Any transaction input referencing a transaction output in a bin, either as a mixin or spend, must also reference all other outputs in that bin. Thus, a real spend cannot be distinguished by age from the other mixin outputs in the bin. Additionally, binned mixin sampling ensures that all the outputs in a bin cannot be deduced as spent until the last unspent output in the bin is spent, preventing deduction attacks from reducing the effective-untraceability of an output to less than the bin size. Algorithm 2 specifies our binned mixin sampling procedure.

The simplest way to assign outputs to bins would be to group outputs according to their position within a block; however a malicious miner could, at no cost, choose which outputs are assigned to which bins harming untraceability. Instead we shuffle the assignment of outputs to bins using the block header of the block containing those outputs. Thus, prior to mining a block, a miner cannot know which outputs in that block will be assigned to which bin. To ensure that a spendable output will always be assigned to a bin, any set of outputs that have been confirmed by ten blocks and are not members of a complete bin are merged into the last complete bin. Due to the privacy risks of using coinbase outputs as mixins, our bin assignment scheme could be trivially modified to prevent non-coinbase outputs from being binned with coinbase outputs. Algorithm 7 (in the Appendix) describes our bin assignment procedure.

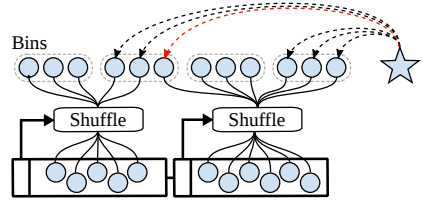

**Fig. 13.** Binned mixin sampling, an input spends an output (red dotted line) and references outputs in two bins.

**Table 4.** Min-untraceability for different bin sizes and mixins. Mixins is the total number of mixins referenced and bin size is the number of outputs per bin. $\varepsilon$ is the maximum percent error

| | | | | Min-untraceability | | | |
|---|---|---|---|---|---|---|---|
| | Bin size | $\varepsilon =$ | 0% | 25% | 50% | 75% | 100% |
| 5 mixins | 1 | | 6.00 | 5.43 | 4.33 | 2.43 | 1.00 |
| 5 mixins | 2 | | 6.00 | 5.18 | 4.00 | 2.67 | 2.00 |
| 5 mixins | 3 | | 6.00 | 5.16 | 4.20 | 3.35 | 3.00 |
| 7 mixins | 1 | | 8.00 | 7.38 | 6.09 | 3.43 | 1.00 |
| 7 mixins | 2 | | 8.00 | 7.02 | 5.43 | 3.26 | 2.00 |
| 7 mixins | 4 | | 8.00 | 6.88 | 5.60 | 4.47 | 4.00 |
| 8 mixins | 1 | | 9.00 | 8.36 | 7.00 | 4.00 | 1.00 |
| 8 mixins | 3 | | 9.00 | 7.76 | 6.00 | 4.00 | 3.00 |

**Choosing the parameters for binned mixin sampling**

If only one bin is referenced, an active attacker could trace a newly created output by broadcasting many attacker controlled outputs. The attack is successful if the targeted output shares a bin with only attacker outputs. Since the number of bins referenced is one when the targeted output is spent, the spending transaction input can only use mixin outputs from that one bin. Thus all the mixin outputs used by the targeted output will be attacker controlled. To prevent such attacks we require that the number of bins referenced be two or greater.

Another benefit of a larger number of bins is that it provides a larger potential range of referenced output ages, thus obfuscating the exact confirmation time of the spent output. Since we must reference at least two bins and we do not wish to require an onerous number of mixins, we consider bin sizes of four or less; the number of mixins required is (number of bins×bin size−1), thus a bin size of five would require all users to use at least nine mixins. Monero as of the Sept. 15, 2017 hard fork requires at least four mixins.

As this countermeasure is designed to resist worst case attacks we introduce a measure of effective-untraceability, called min-untraceability, bounding untraceability under worst case mixin sampling. By worst case assumptions we mean that we choose inputs that reference outputs with ages that maximize the difference between the spend-time distribution and the mixin sampling distribution, thus minimizing the effective-untraceability (our definition of effective-untraceability is given in Section 4.1). Put another way, min-untraceability is the minimum possible effective untraceability given a spend-time and mixin sampling distribution. To represent the degree to which the mixin sampling distribution fails to successfully model the spend-time distribution we parameterize our analysis by the maximum percent error, $\varepsilon$.

We now provide a formal definition of the maximum percent error $\varepsilon$ as used by min-untraceability. Denote the spend-time distribution as $D_S(x)$ and the mixin sampling distribution as $D_M(x)$ where $x$ is the the age of an output. Let $x_{max}$ and $x_{min}$ be output ages which maximize and minimize the ratios

$$r_{max} = \max_{\forall x}\left(\frac{D_S(x)}{D_M(x)}\right), r_{min} = \min_{\forall x}\left(\frac{D_S(x)}{D_M(x)}\right).$$

The difference between $r_{max}$ and $r_{min}$ represents the point of greatest error between the spend-time distribution and the mixin sampling distribution. The maximum percent error $\varepsilon$ is the error between the spend-time distribution and the mixin sampling distribution defined as

$$r_{min} = (1 - \varepsilon), r_{max} = \frac{1}{1 - \varepsilon}.$$

Thus, if $\varepsilon = 0\%$, a temporal analysis attack can never distinguish between the spend-time and mixin sampling distributions, if $\varepsilon = 100\%$ a temporal analysis attack can always distinguish the distributions. A full description of min-untraceability is given in Appendix F.

In Table 4 we compute the min-untraceability for different bin sizes and maximum percent errors using min-untraceability. As shown, increasing the bin size trades off min-untraceability when the maximum percent error is small for increased untraceability when the maximum percent error is large. We argue that a bin size of two is ideal as it maximizes min-untraceability under a small maximum percent error compared to a bin size of three or four, yet still provides an effective-untraceability of 2 against worst-case temporal analysis attacks (attacks which under single mixin sampling would completely trace transactions).

---

**Algorithm 2.** Binned mixin sampling procedure.

SAMPLEBINS(*RealOut, NumMixins, BinSize*)

  NumBins ← (NumMixins + 1)/BinSize;
  RealBin ← MapOutToBin(RealOut);
  MixinVector ← Copy(RealBin);
  **while** |MixinVector| < (NumMixins + 1) **do**
    TxOut ← SampleSingleMixin(RealOut, 1);
    **if** TxOut ∉ MixinVector **then**
      Bin ← MapOutToBin(TxOut);
      MixinVector.appendAll(Bin);

  **return** *sorted*(MixinVector);

---

# 7 Discussion and Recommendations

We have identified two weaknesses in Monero's mixin selection policy, and shown that they pose a significant risk of traceability – especially for early Monero transactions. Next, we discuss how these weaknesses can support investigations into present criminal activity, and also offer suggestions for improving Monero's privacy.

## 7.1 Criminal Uses of Monero

In 2017 there have been three widely publicized instances of criminal activity involving Monero transactions. Our techniques show that Monero is not necessarily a dead end for investigators.

**AlphaBay:** the most prolific darknet market since the Silk Road (operating between December 2014 and July 2017) began accepting Monero deposits in July 2016, partially leading to the large rise in transaction volume. In July 2017, US law enforcement raided an AlphaBay server and seized 12,000 Monero (worth around $500,000) [1]. Assuming the AlphaBay server kept logs generated by the default Monero client, the seized logs could include Monero transactions associated with user withdrawals and deposits, including those prior to 2017.

**Shadow Brokers:** From June 2017 onward, the "Shadow Brokers" offered to accept Monero payments for subscription access to zero-day vulnerabilities and exploit tools. They (mistakenly?) advised their hopeful subscribers to publish their email addresses (hexencoded, but publicly visible) in the Monero blockchain, leading to these transactions being identified [31].

**WannaCry:** Ransomware operators received Bitcoin ransomware payments, to a common address. To launder the Bitcoin ransoms, the operators began ex-changing them for Monero using the Swiss exchange service ShapeShift. The Swiss exchange subsequently announced their cooperation with US law enforcement, and began blacklisting Bitcoin ransoms. However, $36,922 have already been exchanged for Monero [11].

In each of these scenarios, an analyst's goal is to link the criminally-associated transactions to other information, such as accounts at exchanges, which can further their investigation. The analyst starts off having identified several Monero TXOs that belonged to a criminal suspect, and might next ask cooperating exchanges for information about the relatively small number of related transactions referencing that TXO.

A seller at AlphaBay that received a payment directly into an exchange account could clearly be linked this way. Users following a Monero best practice guide, including "How To Use Monero and Not Get Caught" [2], would have known to avoid this by first passing their coins through a wallet on their own computer; however, for transactions made in mid 2016 to early 2017, they might still be traceable through the "deduction" technique.

In the WannaCry and Shadow Brokers scenarios, since the relevant transaction occurred post-RingCT, deduction is mostly likely ineffective. However, analysts could still use the temporal analysis and Guess-Newest heuristic to narrow their search at exchanges or to accumulate probabilistic evidence.

## 7.2 Recommendations

We make the following three recommendations to the Monero community so that privacy can be improved for legitimate uses in the future.

### The mixing sampling distribution should be modified to closer match the real distribution

We have provided evidence that the strategy by which Monero mixins are sampled results in a different time distribution than real spends, significantly undermining the untraceability mechanism. To correct this problem, the mixins should ideally be sampled in such a way that the resulting time distributions match.

A report from Monero Research Labs cited the difficulty of frequently tuning parameters based on data collection (especially since the data collection mechanism itself becomes a potential target for an attacker hoping to alter the parameters) [16]. Fortunately, we provide preliminary evidence that the distribution of "spend-times" changes very little over time. Hence we recommend a sampling procedure based on a model of spending times derived from blockchain data, as discussed in Section 6.1.

### Avoid including publicly deanonymized transaction outputs as mixins

We have empirically shown the harmful effect of publicly deanonymized (i.e. 0-mixin) transactions on the privacy of other users. Since non-privacy-conscious users may make 0-mixin transactions to reduce fees, Monero had instituted a 2-mixin minimum, and recently increased this to 4. However, even 4+mixin transactions may be publicly deanonymized; in particular, as discussed in Section 5.1, mining pools have a legitimate interest in forgoing anonymity by publicly announcing their blocks and transactions for the sake of accountability. Thus, we propose that Monero develop a convention for flagging such transactions as "public," so that other users do not include them as mixins.

### Monero users should be warned that their prior transactions are likely vulnerable to tracing analysis

A significant fraction (91%) of non-RingCT Monero transactions with one or more mixins are deducible (i.e., contain at least one deducible mixin), and therefore can be conclusively traced. Furthermore, we estimate that among all transaction inputs so far, the Guess-Newest heuristic can be used to identify the correct mixin with 80% accuracy. Even after accounting for publicly deanonymized transactions such as pool payouts, we find that at least a few hundred transactions per day in mid 2016 and more than a thousand transactions per day from September 2016 through January 2017 would be vulnerable. Furthermore, we estimate that at most a quarter of these can be attributed to illicit marketplaces like AlphaBay. These users might have incorrectly assumed that Monero provided much higher privacy, especially for transactions taking place in late 2016. Because many transactions on AlphaBay are criminal offenses, with statutes of limitations that will not expire for many years (if ever), these users remain at risk of deanonymization attacks. We stress that illicit businesses tend to be early adopters of new technology, but there exist many legitimate reasons to use privacy-centric cryptocurrencies (e.g., a journalist protecting her sources). While such scenarios are less visible, their users face the same risk of deanonymization.

Towards fulfilling this recommendation, we released an initial draft of this paper to the Monero community. We believe it has been in the best interest of Monero users that we offered this warning as soon as possible, even before countermeasures have been deployed. One reason for our decision is that the data from the Monero blockchain is public and widely replicated, and thus delaying the release would not mitigate post-hoc analysis, which can be carried out at any future time. Second, countermeasures in future versions of the Monero client will not affect the vulnerability of transactions occurring between the time of our publication and the deployment of such future versions.

Complementing this paper, we have launched a block explorer (https://monerolink.com), which displays the linkages between transactions inferred using our techniques. We recommend additionally developing a wallet tool that users can run locally to determine whether their previous transactions are vulnerable.

**Acknowledgements**

We would like to thank the anonymous reviewers and our shepherd Aaron Johnson for their helpful comments and feedback, Mayank Varia for helpful discussions on binned mixins and the MIT PRIMES program. We also thank the Monero community, especially Justin Ehrenhofer, for their feedback on earlier drafts of this paper.

Andrew Miller is supported in part by the Initiative for Cryptocurrencies and Contracts, and is a board member of the Zcash Foundation. Arvind Narayanan is supported by NSF Awards CNS-1421689 and CNS-1651938. This material is based upon work supported by the National Science Foundation under Grant No. 1350733, 1347525, 1414119. Kyle Soska and Nicolas Christin's contributions are based on research sponsored by DHS Office of Science and Technology under agreement number FA8750-17-2-0188. Partial support for this work was provided by the MassTech Collaborative Research Matching Grant Program, as well as the several commercial partners of the Massachusetts Open Cloud. The U.S. Government is authorized to reproduce and distribute reprints for Governmental purposes notwithstanding any copyright notation thereon.

# A Deducibility Attack as a SAT problem

A limitation of the query-based algorithm is that it can only deduce a spend when previous iterations ruled out all other mixins. However, more complex "puzzles" are conceivable where we cannot rule out mixins because they have been provably been spent, but because they must have been spent by another input (cf. Figure 14). We formalize this combinatorial puzzle as a SAT problem and solve it using the Sat4j SAT solver.

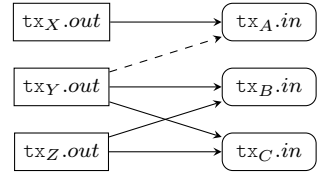

**Fig. 14.** Since $\text{tx}_Y.out$ must have been spent by either $\text{tx}_B.in$ or $\text{tx}_C.in$, $\text{tx}_Y.out$ must have been spent by $\text{tx}_A.in$.

A SAT problem $p$ is a boolean formula containing $n$ propositional variables that yield a mapping $\delta$ between inputs and outputs. Each variable $r_{xy} \in \delta$ denotes a single reference from an input $x$ to an output $y$. Let $\delta(\hat{x})$ give us all existing $y$ values for $r_{\hat{x}y}$, and vice versa. We can then specify restrictions on the possible relationships between inputs and outputs.

– Every input can spend any of the referenced outputs:

$$\bigwedge_{x=1}^{|X|} \bigvee_{y \in \delta(x)} r_{xy}$$

– An input can only spend a single output:

$$\bigwedge_{x=1}^{|X|} \bigwedge_{i=1}^{|\delta(x)|-1} \bigwedge_{j=i+1}^{|\delta(x)|} (\neg r_{x\delta(x)_i} \lor \neg r_{x\delta(x)_j})$$

– Similarly, every output can only be spent by a single input:

$$\bigwedge_{y=1}^{|Y|} \bigwedge_{i=1}^{|\delta(y)|-1} \bigwedge_{j=i+1}^{|\delta(y)|} (\neg r_{\delta(y)_i y} \lor \neg r_{\delta(y)_j y})$$

Since inputs can only spend outputs of the same value, the SAT problem can be solved individually for each denomination. Assignments of propositional variables that are `true` in all solutions of $p$ correspond to true spends, as it reflects the only possible way to spend an output (e.g., $\text{tx}_A.in$ always spends $\text{tx}_X.out$ in Figure 14).

Solving the SAT problem using Sat4j yields an additional 5149 deanonymized spends across 1157 different denominations.

# B Detailed Description of our Models of Monero Sampling Routines

In Algorithms 3, 4, 5, 6 we give pseudocode for the mixin selection procedures used in our simulation (Section 4). The changes between each successive version are highlighted in blue. We note that these are simplified models of the mixin sampling behavior in the Monero client, in particular they elide over edge cases that avoid spending "locked" coins that have recently been mined. The full code listing of the Monero client can be found on the Monero git repository.[2]

---

**Algorithm 3.** SAMPLEMIXINS(*RealOut*, *NumMixins*)
**[vPre0.9.0]**

Let TopGIdx be the index of the most recent transaction output with denomination *RealOut.amount*;
BaseReqMixCount := $\lfloor (NumMixins + 1) \times 1.5 + 1 \rfloor$;
**while** |MixinVector| < BaseReqMixCount **do**
  $i \leftarrow$ UniformSelect(*0*, TopGIdx);
  **if** $i \notin$ MixinVector and $i \neq RealOut.idx$ **then**
    MixinVector.append($i$);

Let FinalVector be a uniform random choice of
  *NumMixins* elements from MixinVector;
**return** *sorted*(FinalVector + [RealOut.idx]);

---

**Algorithm 4.** SAMPLEMIXINS(*RealOut*, *NumMixins*)
**[v0.9.0]**

Let TopGIdx be the index of the most recent transaction output with denomination *RealOut.amount*;
BaseReqMixCount := $\lfloor (NumMixins + 1) \times 1.5 + 1 \rfloor$;
MixinVector := [];
**while** |MixinVector| < BaseReqMixCount **do**
  $i \leftarrow$ TriangleSelect(*0*, TopGIdx);
  **if** $i \notin$ MixinVector and $i \neq RealOut.idx$ **then**
    MixinVector.append($i$);

Let FinalVector be a uniform random choice of
  *NumMixins* elements from MixinVector;
**return** *sorted*(FinalVector + [RealOut.idx]);

---

**Algorithm 5.** SAMPLEMIXINS(*RealOut*, *NumMixins*)
**[v0.10.1]**

Let TopGIdx be the index of the most recent transaction output with denomination *RealOut.amount*;
BaseReqMixCount := $\lfloor (NumMixins + 1) \times 1.5 + 1 \rfloor$;
Let RecentGIdx be the index of the most recent transaction output prior to 5 days ago with denomination *RealOut.amount*;
BaseReqRecentCount := MAX(1, MIN(     TopGIdx - RecentGIdx + 1,        BaseReqMixCount × RecentRatio));
**if** $RealOut.idx \geq$ RecentGIdx **then**
  BaseReqRecentCount -= 1
MixinVector := [];
**while** |MixinVector| < BaseReqRecentCount **do**
  $i \leftarrow$ UniformSelect(RecentGIdx, TopGIdx);
  **if** $i \notin$ MixinVector and $i \neq RealOut.idx$ **then**
    MixinVector.append($i$);

**while** |MixinVector| < BaseReqMixCount **do**
  $i \leftarrow$ TriangleSelect(*0*, TopGIdx);
  **if** $i \notin$ MixinVector and $i \neq RealOut.idx$ **then**
    MixinVector.append($i$);

Let FinalVector be a uniform random choice of
  *NumMixins* elements from MixinVector;
**return** *sorted*(FinalVector + [RealOut.idx]);

---

# C An Analysis of Bytecoin

The cryptocurrency Bytecoin was an early implementation of the Cryptonote protocol. As Monero is based upon Bytecoin's codebase, Bytecoin's mixin sampling procedure shares the same weaknesses. We run the mixin sudoku algorithm on transaction data extracted from the Bytecoin blockchain and show the results in Table 5. Overall, we are able to deduce the real spent for 29% of all inputs with more than one mixin. Of those that include only 1 mixin, we can deduce 56% of inputs. We attribute the lower total success rate to a discrepancy between the number inputs to outputs in Bytecoin, suggesting that there exist a lot of unspent outputs from which significantly fewer inputs can choose.

In Table 6 we show the percentage of inputs where guessing the most recent output yields the true spend. With an accuracy of 97.54% the Guess-Newest heuristic proves to be very effective.

---

**2** https://github.com/monero-project/monero/blob/v0.9.0/src/wallet/wallet2.h#L570 and https://github.com/monero-project/monero/blob/v0.10.0/src/wallet/wallet2.cpp#L3605

**Algorithm 6.** SAMPLEMIXINS(*RealOut*, *NumMixins*) **[v0.11.0]**

Let TopGIdx be the index of the most recent transaction output;

BaseReqMixCount := $\lfloor (NumMixins + 1) \times 1.5 + 1 \rfloor$;

Let RecentGIdx be the index of the most recent transaction output prior to 1.8 days ago;

BaseReqRecentCount := MAX(1, MIN( TopGIdx - RecentGIdx + 1, BaseReqMixCount × RecentRatio));

**if** $RealOut.idx \geq$ RecentGIdx **then**
 ⌊ BaseReqRecentCount -= 1

MixinVector := [];

**while** |MixinVector| < BaseReqRecentCount **do**
 $i \leftarrow$ TriangleSelect(RecentGIdx, TopGIdx);
 **if** $i \notin$ MixinVector **and** $i \neq RealOut.idx$ **then**
  ⌊ MixinVector.append($i$);

**while** |MixinVector| < BaseReqMixCount **do**
 $i \leftarrow$ TriangleSelect(0, TopGIdx);
 **if** $i \notin$ MixinVector **and** $i \neq RealOut.idx$ **then**
  ⌊ MixinVector.append($i$);

Let FinalVector be a uniform random choice of *NumMixins* elements from MixinVector;

**return** *sorted*(FinalVector + [RealOut.idx]);

**Table 5.** Bytecoin transaction inputs (with 1 or more mixins, at least 1000 TXOs available) where the real input can be deduced.

|  | Total | Deducible | (%) |
|---|---|---|---|
| 1 mixins | 4192272 | 2338746 | (55.79) |
| 2 mixins | 813375 | 264610 | (32.53) |
| 3 mixins | 1243428 | 207540 | (16.69) |
| 4 mixins | 2450891 | 249618 | (10.18) |
| 5 mixins | 405140 | 25666 | (6.34) |
| 6 mixins | 1250627 | 51755 | (4.14) |
| 7 mixins | 158753 | 4198 | (2.64) |
| 8 mixins | 76530 | 539 | (0.70) |
| 9 mixins | 62714 | 226 | (0.36) |
| 10 mixins | 204725 | 197 | (0.10) |
| Total | 10858455 | 3143095 | (28.95) |

**Table 6.** Percentage of deducible Bytecoin transaction inputs where the real input is the "newest" input.

|  | Deducible | Newest | (%) |
|---|---|---|---|
| 1 mixins | 2338746 | 2311078 | (98.82) |
| 2 mixins | 264610 | 249000 | (94.10) |
| 3 mixins | 207540 | 193453 | (93.21) |
| 4 mixins | 249618 | 245236 | (98.24) |
| 5 mixins | 25666 | 19027 | (74.13) |
| 6 mixins | 51755 | 44243 | (85.49) |
| 7 mixins | 4198 | 3011 | (71.72) |
| 8 mixins | 539 | 378 | (70.13) |
| 9 mixins | 226 | 149 | (65.93) |
| 10+ mixins | 197 | 128 | (64.97) |
| Total | 3143095 | 3065703 | (97.54) |

# D Guessing Entropy Untraceability

We construct a measure of the guessing entropy of the untraceability of a transaction input which we denote Ge. We compute this value as

$$\text{Ge} = \sum_{0 \leq k \leq M} k \cdot p_k$$

where $p = p_0 \geq p_1 \geq \cdots \geq p_M$ are the probabilities that each referenced output is the spent output, and $M$ is the number of mixins outputs referenced by the transaction input for $M + 1$ total referenced outputs. Put another way Ge is the expected number of guesses beyond the first guess to correctly guess the spent output. For instance if $M = 3$ and each referenced output is equally likely ($p_i = \frac{1}{4}$) then the expected number of guesses would be Ge = $0 \cdot \frac{1}{4} + 1 \cdot \frac{1}{4} + 2 \cdot \frac{1}{4} + 3 \cdot \frac{1}{4} = 3/2$.

The effective-untraceability, i.e. effective-anonymity set size, of a transaction input is $(1 + 2 \cdot \text{Ge})$. If all referenced outputs of the transaction input are equally likely to be the real spend the effective-untraceability for that transaction input would be $M + 1$. However if some outputs are more likely than others the guessing entropy, and thus the effective-untraceability, decreases to reflect this loss of untraceability. For instance consider a trans-action input with $M = 3$ mixins that has a guessing entropy of $1/2$ rather than the ideal guessing entropy of $3/2$. Such a transaction input would have a effective-untraceability of 2 or the equivalent untraceability to an ideal transaction input with only $M = 1$ mixins.

# E Bin Assignment Scheme

---

**Algorithm 7.** Our bin assignment scheme for binned mixin sampling. ASSIGNBINS(*Blockchain*, *BinSize*)

TxOutVector := [];
Bins := [];
**for** $h \leftarrow 0$, $h <$ Height(Blockchain), $h \leftarrow h + 1$ **do**
    Block $\leftarrow$ Blockchain[$h$];
    $r \leftarrow$ Hash(Block.Header);
    ShuffledTxOuts $\leftarrow$ Shuffle($r$, Block.TxOuts);
    TxOutVector.appendAll(ShuffledTxOuts);

    **while** |TxOutVector| > (2 × BinSize − 1) **do**
        Bin $\leftarrow$ TxOutVector.pop(BinSize);
        Bins.append(Bin);

Bins.append(TxOutVector);
**return** Bins;

---

# F Minimum Untraceability

Let the spend-time distribution be denoted as $D_S(x)$ and the mixin sampling distribution be denoted as $D_M(x)$ where $x$ is the age of an output. If these two distributions are known, an attacker can use temporal analysis to answer the question if we observe a transaction input with the following ages of referenced outputs $X = x_0, x_1, x_2, \cdots, x_m$, what is the probability $p_i = P(x_i|X)$ that the referenced input at time $x_i$ is the real spend.

$D_M(x)$ and $D_S(x)$ allows us to compute $P(X|x_i)$, the probability of $X$ being selected given that $x_i$ is the spent output

$$P(X|x_i) = \prod_{0 \leq q \neq i \leq m} D_M(x_q),$$

the probability that $x_i$ would be selected from the spend-time distribution

$$P(x_i) = D_S(x_i),$$

and the probability that $X$ is referenced by the input

$$P(X) = \sum_{0 \leq p \leq m} (D_S(x_p) \prod_{0 \leq j \neq p \leq m} D_M(x_j)).$$

Using Bayes' theorem we can now compute the probability that $x_i$ is the spent output

$$P(x_i|X) = \frac{P(x_i)P(X|x_i)}{P(X)}$$

$$= \frac{D_S(x_i) \prod_{0 \leq j \neq i \leq m} D_M(x_j)}{\sum_{0 \leq p \leq m} \left( D_S(x_p) \prod_{0 \leq q \neq p \leq m} D_M(x_q) \right)}$$

$$= \frac{\frac{D_S(x_i)}{D_M(x_i)}}{\sum_{0 \leq p \leq m} \frac{D_S(x_p)}{D_M(x_p)}}$$

Let $r_x = \frac{D_S(x)}{D_M(x)}$ be the ratio between the spend-time distribution and the mixin sampling distribution for some age $x$. We can reformulate $P(x_i|X)$ in terms of $r_x$

$$p_i = P(x_i|X) = \frac{r_i}{\sum_{0 \leq p \leq m} (r_p)}.$$

Plugging in our expression for $p_i = P(x_i|X)$ and sorting the probabilities so that $p_0 \geq p_1 \geq \cdots \geq p_m$ we can compute the guessing entropy Ge as a function of the ratio $r_x$

$$\text{Ge} = \sum_{0 \leq k \leq m} k \cdot \frac{r_k}{\sum_{0 \leq p \leq m} (r_p)},$$

We can now use this definition to measure temporal analysis. For instance, if $r_x = 1.0$ for an output of age $x$, this output provides no information for temporal analysis. On the other hand we can eliminate as mixins any referenced output ages $x$ such that $r_x = 0.0$. More generally if the ratio of two referenced outputs $x_i, x_j$ is $r_i > r_j$ then $x_i$ has a higher probability of being the spent output than $x_j$, i.e. $p_i > p_j$.

To perform a worst case analysis we need to find the ages that minimize the guessing entropy. Let $x_{max}$ and $x_{min}$ be output ages which maximize and minimize the ratio $r$ where

$$r_{max} = \max_{\forall x}(\frac{D_S(x)}{D_M(x)}), r_{min} = \min_{\forall x}(\frac{D_S(x)}{D_M(x)}).$$

The difference between $r_{max}$ and $r_{min}$ represents the point of greatest error between the spend-time distribution and the mixin sampling distribution. We denote the maximum percent error between the spend-time distribution and the mixin sampling distribution as $\varepsilon$ such that

$$r_{min} = (1 - \varepsilon), r_{max} = \frac{1}{1 - \varepsilon},$$

and

$$\frac{r_{max}}{r_{min}} = \frac{1}{(1 - \varepsilon)^2}.$$

The most vulnerable possible set of referenced outputs to temporal analysis is a single output whose age is $x_{max}$ (most likely to be the real spend) and where all other referenced outputs have ages equal to $x_{min}$ (most likely to be mixins). Such a set of transaction inputs yields $\text{Ge}_{min}$, the minimum possible guessing entropy of the untraceability of a transaction input.

$$\text{Ge}_{min} = \frac{0 \cdot r_{max} + \sum_{1 \leq k \leq m} k \cdot r_{min}}{r_{max} + \sum_{1 \leq k \leq m} (k \cdot r_{min})} = \frac{\frac{1}{2}m(m+1)}{\frac{r_{max}}{r_{min}} + m}$$

We can reformulate $\mathrm{Ge}_{min}$ in terms of the maximum percent error of the sampling distribution, $\varepsilon$, and number of mixins, $m$,

$$\mathrm{Ge}_{min}(m, \varepsilon) = \frac{\frac{1}{2} m(m+1)}{\frac{1}{(1-\varepsilon)^2} + m}$$

The above function only gives us $\mathrm{Ge}_{min}$, the guessing entropy for the current mixin sampling procedure. We will now compute it for binned mixins. Binned mixins have two parameters: the size of each bin $s$ and the number of bins $n$. The total number of mixins outputs referenced by a transaction input using mixin bins is $m = (n \cdot s) - 1$. We use $\mathrm{Ge}_{min}(m, \varepsilon)$ to create a function, $\mathrm{BGe}_{min}(s, n, \varepsilon)$, for the minimum guessing entropy for binned mixins.

As each bin contains outputs of the same age, all outputs sharing a bin share the same probability under temporal analysis of being the spent output. Thus, we can treat bins as transaction outputs which require more than one guess to search. We will now formally show that this is the case. Let $p_i$ be the probability the $i$-th bin contains the spent output, then $q = \frac{p_i}{s}$ is the probability that an output in the $i$-th bin is the real spend. This gives us a function for the guessing entropy untraceability of binned mixins:

$$\mathrm{BGe} = \sum_{k=0}^{ns-1} (k \cdot q_k) = q_0 \sum_{j=0}^{s-1} j + q_1 \sum_{j=s}^{2s-1} j + \cdots + q_{n-1} \sum_{j=s \cdot (n-1)}^{(s \cdot n)-1} j$$

$$= \sum_{k=0}^{n-1} q_k \left( \sum_{i=ks}^{(k+1) \cdot s-1} i \right) = \sum_{k=0}^{n-1} \frac{p_k}{s} \left( \frac{s}{2} (2ks + s - 1) \right)$$

$$= \frac{1}{2} \left( 2s \sum_{k=0}^{n-1} (k \cdot p_k) + s - 1 \right) = s \sum_{k=0}^{n-1} (k \cdot p_k) + \frac{s-1}{2}$$

We plug $\mathrm{Ge}_{min}$ into $\mathrm{BGe}$ to create a function for guessing entropy untraceability for binned mixins:

$$\mathrm{BGe}_{min}(s, n, \varepsilon) = s \cdot \mathrm{Ge}_{min}(n-1, \varepsilon) + \frac{s-1}{2}.$$

For bin sizes of $s = 1$ this is just $\mathrm{Ge}_{min}$:

$$\mathrm{BGe}_{min}(s = 1, n = m+1, \varepsilon) = 1 \cdot \mathrm{Ge}_{min}(n-1, \varepsilon) + \frac{1-1}{2}$$

$$= \mathrm{Ge}_{min}(n-1, \varepsilon) = \mathrm{Ge}_{min}(m, \varepsilon).$$

This makes sense as binned mixins with a bin size $s = 1$ are equivalent to the current mixin sampling procedure. The min-untraceability is the effective-untraceability using $\mathrm{BGe}_{min}$ as the guessing entropy.